\documentclass[prb,aps,final,twocolumn,floatfix]{revtex4}
\input epsf
\begin{document}
\title{Theoretical study of the finite temperature spectroscopy in van der
Waals clusters. III Solvated Chromophore as an effective diatomics}
\author{F. Calvo, F. Spiegelman}
\affiliation{Laboratoire de Physique Quantique, IRSAMC, Universit\'e Paul
Sabatier, 118 Route de Narbonne, F31062 Toulouse Cedex, France}
\author{J.-M. Mestdagh}
\affiliation{CEA/DRECAM, CE Saclay, 91191 Gif-sur-Yvette Cedex, France}
\begin{abstract}
The absorption spectroscopy of calcium-doped argon clusters
is described in terms of an effective diatomics molecule
Ca--(Ar$_n$), in the framework of semiclassical vertical
transitions. We show how, upon choosing a suitable reaction
coordinate, the effective finite-temperature equilibrium
properties can be obtained for the ground- and
excited-surfaces from the potential of mean force (PMF). An
extension of the recent multiple range random-walk
method is used to calculate the PMF over continuous intervals
of distances. The absorption spectra calculated using this
single-coordinate description are found to be in good
agreement with the spectra obtained from high-statistics
Monte Carlo data, in various situations. For CaAr$_{13}$, we
compare the performances of two different choices of the
reaction coordinate. For CaAr$_{37}$, the method is seen to
be accurate enough to distinguish between different
low-energy structures. Finally, the idea of casting the
initial many-body problem into a single degree of freedom
problem is tested on the spectroscopy of calcium in bulk
solid argon.
\end{abstract}
\maketitle

\section{Introduction}

Polyatomic molecules can be electronically excited in a
global or local way, depending on the chemical nature of the system, and
on details of the
excitation mechanisms, such as the characteristics or of the
laser pulse. In many systems, the chemical bonds at
various places of the same molecule favor local excitations.
This situation typically occurs in chromophore-doped rare-gas
clusters, where photons in the visible range may excite the
chromophore but not the rare-gas atoms. This
feature is especially useful because it allows a precise
analysis by comparison with the isolated chromophore
spectroscopy. Actually, even
though the chromophore is locally excited, the geometric
distortions and the vibrations of the solvent atoms
act as a perturbation on its spectroscopic properties. From a
theoretical point of view, these many-body effects are
usually treated either with simple
approximations,\cite{yan,wp} or conveniently by simulation.
In the previous papers of this series,\cite{paperI,paperII}
we have presented an alternative
approach to simulation, where the Gaussian theory of
absorption by Wadi and Pollak\cite{wp} was combined with the
superposition approximation\cite{jcpquantum} in the quantum
regime at finite temperature. This method was applied to
CaAr$_n$ clusters of various sizes $n$ in the range $2\leq
n\leq 146$.

In these molecular systems, the calcium atom is energetically
more stable when located on the surface of the argon cluster,
rather than in a fully solvated site. At moderate
temperatures, below the cluster melting point, the calcium
atom gets the ability to glide over the argon
cluster.\cite{paperI} The heterogeneous cluster
CaAr$_n$ can then be thought as an effective diatomic
molecule, where the argon cluster would be replaced by a
large pseudo-atom. Similar ideas have been previously used in
chemical physics or condensed matter physics, as in the
effective interactions between colloids or between polymers,
or in the Girifalco\cite{girifalco} or
Gspann-Vollmar\cite{gspann} interaction potentials between
clusters of carbon or rare-gas atoms, respectively.

In this work, we investigate the possibility of treating the
photoabsorption spectroscopy of CaAr$_n$ clusters in this
effective diatomics picture. For this purpose, we need to
characterize the interaction on the ground- and excited-state
potential energy surfaces in terms of a single reaction
coordinate separating the calcium atom from the argon cluster. The
effective absorption spectrum can then be constructed from
the effective potential curves. The finite-temperature
properties of a many-body system are conveniently expressed
as ensemble averages in the canonical ensembles. It is
possible to treat the reaction coordinate separately, and to
perform some averages on the remaining degrees of freedom.
The next stage consists of averaging over the reaction
coordinate itself using its suitable statistical weight. In
this context, the statistical average on the many-body system
is reduced into a potential of mean force (PMF) calculation
that will appear as a Helmholtz free energy.
Several methods are available to get the computational
solution to this problem,\cite{allen,frenkel} including
umbrella sampling\cite{valleau} or constraint
dynamics,\cite{constdyn} or more recently the multiple-range
random walk algorithm,\cite{wl,fcmp} which we have used in the
present work.

The paper is organized as follows. In the next section, we
describe the general method, and the algorithms used to
calculate the effective properties of the Ca--(Ar$_n$)
cluster as a function of an internal reaction coordinate. We
then apply in Sec.~\ref{sec:app} the method to various
cluster sizes, at several temperatures. In particular, the
choice of the reaction coordinate may not be systematically
obvious, and we show how more information can be obtained
from a carefully chosen coordinate. For a suitable choice,
absorption spectroscopy of calcium in bulk argon can also be
studied, and some results will be presented in section
\ref{sec:app}. We finally summarize and conclude in
Sec.~\ref{sec:ccl}.

\section{Free-energy profiles for ground- and excited-states}
\label{sec:fep}

Our system is a CaAr$_n$ cluster, described using the
Cartesian atomic coordinates ${\bf R}=\{x_i,y_i,z_i\}$, where
subscript 0 will be used for the calcium atom. We first
present the main ideas and approximations of the method.

The photoabsorption spectrum of the cluster is calculated in
a semiclassical way by assuming vertical transitions (Condon
approximation) between the ground state surface $V_0({\bf
R})$ and several excited-state surfaces $V_k({\bf R})$. At
each configuration ${\bf R}$, the absorption cross section
$\sigma_k({\bf R})$ is proportional by the square of the transition
dipole moment at this point, scaled by the transition energy.
Excitation from the ground state
surface thermalized at temperature $T$ leads to the
unnormalized absorption intensity ${\cal I}(\omega)$ given by
the sum over excited states, ${\cal I}(\omega)=\sum_k {\cal
I}_k(\omega)$. The intensity of absorption from the ground
state $0$ to state $k$ is

\begin{equation}
{\cal I}_k(\omega) = \frac{1}{Z}\int \delta\left\{
\hbar\omega - [V_k({\bf R}) -V_0({\bf R})]\right\}
\sigma_k({\bf R}) e^{-V_0({\bf R})/k_BT} d{\bf R},
\label{eq:igen}
\end{equation}
where the partition function $Z(T)$ is calculated on the ground state surface.
For each excited surface $k$, and up to a constant factor, ${\cal I}_k(\omega)$
can be written as a canonical average on $V_0$:
\begin{equation}
{\cal I}_k(\omega) = \langle A_k(\omega,{\bf R})\rangle,
\label{eq:ica}
\end{equation}
with the notations
\begin{equation}
A_k(\omega,{\bf R}) =
\sigma_k({\bf R}) \delta\{ \hbar\omega - [V_k({\bf R})
-V_0({\bf R})]\} ; \label{eq:defa}
\end{equation}
and
\begin{equation}
\langle O \rangle = \frac{1}{Z}\int O({\bf R})
e^{-V_0({\bf R})/k_BT} d{\bf R},
\label{eq:lar}
\end{equation}
for any observable $O({\bf R})$.
In the following, and for clarity reasons, we shall drop the
subscript $k$ indicating the excited state. The single
excited surface will be denoted $V^*$.

Let now assume that a reaction coordinate or an order
parameter $\xi({\bf R})$ can be defined, which characterizes
the overall location of ground state calcium with respect to
the argon atoms. Possible expressions for $\xi({\bf R})$ will
be discussed at the end of the present section. At a given
temperature $T$, calcium has a probability $p(\xi_0)$ of
residing at coordinate $\xi_0$ given by the canonical average
\begin{equation}
p(\xi_0) = \langle \delta[\xi({\bf R})-\xi_0] \rangle.
\label{eq:jmm}
\end{equation}
This defines a potential of mean force (PMF) $W(\xi_0)$
according to:
\begin{equation}
W(\xi_0) = -k_BT \ln p(\xi_0). \label{eq:pmf}
\end{equation}
The absolute value of the PMF is meaningful only if the probability
distribution $p$ is normalized. It can be arbitrarily shifted by any
additive term $W_0$, provided that a factor $e^{W_0/k_BT}$ is incorporated
in the calculated observables.

Building upon these definitions, we can introduce the partial average $\bar
O(\xi_0)$ of observable $O$ by restriction on the coordinate
$\xi$:
\begin{equation}
\bar O(\xi_0)=\frac{\int O({\bf R})\delta[\xi({\bf
R})-\xi_0]\exp[-V_0({\bf R}) /k_BT] d{\bf R}}
{\int \delta[\xi({\bf R})-\xi_0]\exp[-V_0({\bf R}) /k_BT] d{\bf R}}.
\label{eq:arestr0}
\end{equation}
$\bar O$ can also be written
\begin{equation}
\bar O(\xi_0)=\frac{\int O({\bf R})\delta[\xi({\bf
R})-\xi_0]\exp[-V_0({\bf R}) /k_BT] d{\bf R}}{Z p(\xi_0)},
\label{eq:arestr}
\end{equation}
where $Z p(\xi_0)$ plays the role of a restricted partition
function. By definitions of $p$ and $W$, Eqn.~(\ref{eq:jmm}) and
(\ref{eq:pmf}), the global thermal average $\langle
A(\omega,{\bf R}) \rangle$ over the whole configuration space
is given by the one-dimensional average over the coordinate
$\xi$:
\begin{equation}
\langle A(\omega,{\bf R})\rangle = \int \bar A(\xi)
e^{-W(\xi)/k_BT} d\xi. \label{eq:aavre}
\end{equation}
Importantly, Eq.~(\ref{eq:pmf}) indicates that the PMF
$W(\xi)$ plays the role of a Helmholtz free energy term
corresponding to the deformation of the system along the
single coordinate $\xi$. We have thus mapped the initial
many-body problem into a simpler one-dimensional problem,
where the initial ground-state potential has been replaced by
the PMF, and the instantaneous observable by its partial
overage over all other, unrestricted degrees of freedom.

In the present context, further assumptions are
needed in order to get the full picture of an
effective diatomics to describe the CaAr$_n$ cluster.
>From expressions similar to Eq.~(\ref{eq:arestr}) effective
surfaces $\bar V_0(\xi)$ and $\bar V^*(\xi)$, as well as
effective cross sections $\bar\sigma(\xi)$, can be constructed.
The absorption intensity is then approximated as
\begin{eqnarray}
{\cal I}(\omega) &\approx& \bar {\cal I}(\omega) \nonumber \\
&=& \int \delta\{\hbar\omega - [\bar V^*(\xi)-\bar V_0(\xi)]\}\bar\sigma(\xi)
e^{-W(\xi)/k_BT}d\xi. \label{eq:approx1}
\end{eqnarray}
A further approximation can be done, which will be
checked later. We replace the function $\exp(-W/k_BT)$ by a
Boltzmann weight over the effective ground-state surface
$\bar V_0$. Hence
\begin{equation}
{\cal I}(\omega) = C \int \delta\{\hbar\omega - [\bar V^*(\xi)
-\bar V_0(\xi)]\}\bar\sigma(\xi) e^{-\bar V_0(\xi)/k_BT}d\xi,
\label{eq:approx2}
\end{equation}
where the proportionality constant $C$ accounts for the normalization of
$\bar V_0$:
\begin{equation}
C^{-1} = \int e^{-\beta \bar V_0(\xi)/k_BT}d\xi.
\label{eq:cmu}
\end{equation}
In the following, we will only consider normalized absorption spectra, and
$C$ will be dropped.

Using this extra approximation, Eq.~(\ref{eq:approx2}) thus obtained is the
complete analogue of Eq.~(\ref{eq:igen})
for a single-coordinate system. In order to use it in a practical situation,
we need to calculate all effective quantities over a continuous range of the
reaction coordinate $\xi$, starting with the free-energy profiles $\bar V_0
(\xi)$ and $\bar V^*(\xi)$. For this purpose we use the multiple range random
walk method of Wang and Landau\cite{wl} recently extended to the calculation
of potentials of mean force and free-energies.\cite{fcmp}

Briefly, we introduce a function $g(\xi)$ initially set to 1 in the range of
accessible values of $\xi$, and we set $s(\xi)=\ln g(\xi)$. A Monte Carlo
simulation is carried out using the following Metropolis acceptance
rule between the old ${\bf R}_{\rm old}$ and new ${\bf R}_{\rm new}$
configurations:\cite{wl,fcmp}
\begin{equation}
{\rm acc}({\bf R}_{\rm old}\to{\bf R}_{\rm new}) = \min[1,\exp(-\Delta
F/k_BT)],
\label{eq:metropolis}
\end{equation}
with $\Delta F = F({\bf R}_{\rm new})-F({\bf R}_{\rm old})$
and the (Landau) free energy $F({\bf R}) = V_0({\bf R}) -k_BT
s[\xi({\bf R})]$. After the new
configuration ${\bf R}_{\rm new}$ is visited, $s$ is updated:
$s[\xi( {\bf R}_{\rm new})]\to s[\xi({\bf R}_{\rm new})] +
\ln f$, or equivalently, $g$ is multiplied by $f$.
Here $f$ is a fixed quantity, initially set to 2--2.5. After
a large number of MC steps, a new iteration $m$ starts where
$\alpha=\ln f$ is reduced by taking $\alpha_{m+1}=
\alpha_m/2$. In this algorithm, the function $\Gamma(\xi) =
-k_BT s(\xi)$ smoothly converges to the potential of mean
force $W(\xi)$, up to an additive factor.

Once it has been calculated, the PMF gives also access to the
restricted averages $\bar A$ over a much wider range of $\xi$
than normally accessible. For this purpose each configuration in the Monte
Carlo simulation is reweighted by the factor $\exp\{-W[\xi({\bf R})]\}$. One
should then get a uniform probability distribution for $\xi$. The
flatness of this distribution can be used as a check of the
PMF computed using the Wang-Landau method.\cite{wl} The
effective averaged observables $\bar A$ are then obtained
with the usual reweighting formulas.\cite{allen}

We now turn to the choice of the reaction coordinate
$\xi({\bf R})$ in concern with absorption spectroscopy. A
suitable coordinate should have a clear geometrical meaning.
A straightforward definition is the ``local'' distance $d({\bf
R})$ between calcium and the closest argon atom:
\begin{equation}
d({\bf R}) = \min_{i\geq 1} \| \vec r_i - \vec r_0 \|,
\label{eq:ddef}
\end{equation}
where $\vec r_i$ denotes the position vector of atom $i$.
This definition has the advantage that it reduces to the usual
interatomic distance for the true diatomic molecule CaAr.
Furthermore, this choice does not suffer any problem in a MC
framework, in spite of the discontinuities that definition
(\ref{eq:ddef}) introduces in the Jacobian. In contrast, this
would not be the case in a molecular dynamics simulation and
an alternative better choice would be the distance between
the chromophore and the center of mass (com) of the argon
cluster:
\begin{equation}
d_{\rm com}({\bf R}) = \left\| \vec r_0 - \frac{1}{n}\sum_{i=1}^n \vec r_i
\right\|.
\label{eq:dcdm}
\end{equation}
Let us notice that the latter reaction coordinate is not practical in several
cases, such as calcium in bulk argon. However, it can
sometimes bring extra information with respect to the choice
of $d$ above, Eq.~(\ref{eq:ddef}), as will appear in the next section.

\section{Applications to C\lowercase{a}@A\lowercase{r}}
\label{sec:app}

We have applied the ideas developed in the previous section
to two CaAr$_n$ clusters, namely CaAr$_{13}$ and CaAr$_{37}$.
The $4s^2$ ground state potential is modelled by
simple pairwise terms, and the $4s4p$ excited states are
modelled using a Diatomic-In-Molecules (DIM)
Hamiltonian. All details about the potentials and the
parameterization are given in Ref.~\onlinecite{epjd}.

For all systems investigated here, the reference absorption
spectra were obtained from classical parallel tempering Monte
Carlo, using $10^6$ cycles after $2\times 10^5$ initial
cycles discarded for equilibration. For CaAr$_{37}$, we did
not use parallel tempering because only the local properties
of the isomers were needed. The potential of mean force was
computed from the multiple range random walk algorithm using
20 iterations of each $10^6$ cycles following $2\times 10^5$
thermalization cycles. The parameter $f$ was initially taken
as 2.5, and the number of bins in the histograms of the
reaction coordinate was set to 1000.

CaAr$_{13}$ is an obvious choice for testing the above
methods, because the Ar$_{13}$ is icosahedral, hence nearly spherical,
and because in this system calcium occupies a
capping site of the icosahedral cluster. This is a
virtual support to the effective diatomic picture. As seen
in our previous papers,\cite{paperI,paperII} the calcium atom
can jump into the icosahedral shell at moderate temperatures
$T\sim 30$~K, but the spectroscopic signature of these
isomerizations is rather weak.

\begin{figure}[htb]
\vbox to 7.6cm{
\includegraphics{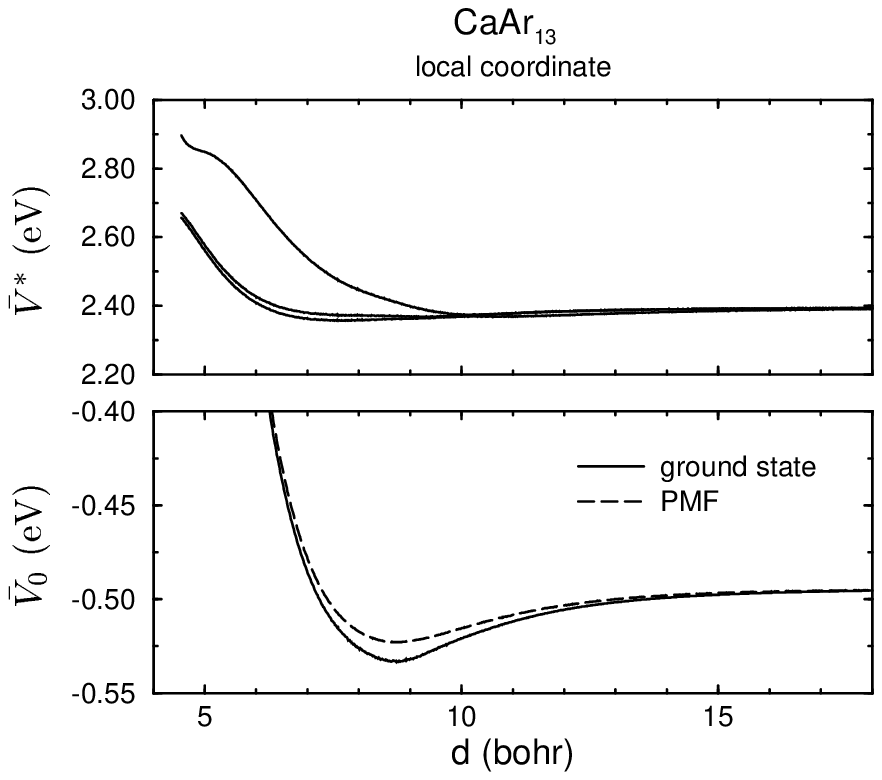}
\vfill}
\caption{Effective potential surfaces of CaAr$_{13}$ at $T=30$~K, as function
of the local distance coordinate $d$. $V_0$ and $V^*$ are the ground- and
excited-states surfaces, respectively. On the lower panel, the potential of
mean force (PMF) is also represented, after shifting its asymptotic value to
the ground-state potential.}
\label{fig:pot13}
\end{figure}
\begin{figure}[htb]
\vbox to 6.6cm{
\includegraphics{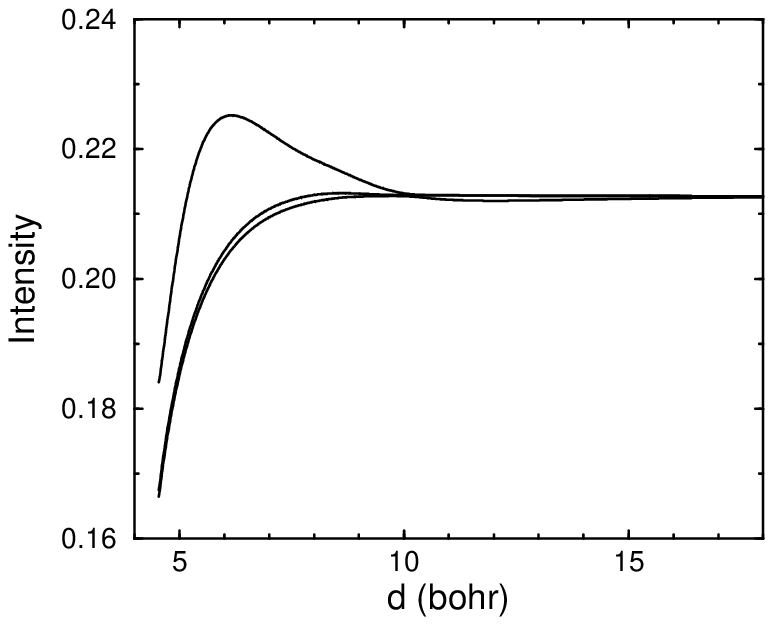}
\vfill}
\caption{Effective absorption intensities of CaAr$_{13}$ at $T=30$~K as
function of the local coordinate $d$.}
\label{fig:int13}
\end{figure}

In Fig.~\ref{fig:pot13} we have represented the effective
potential energy curves (ground and excited states) of
CaAr$_{13}$ at $T=30$~K, computed using the local distance $d$
between calcium and argon defined by Eq.~(\ref{eq:ddef}). We
have superimposed the (shifted) potential of mean force on
the ground state effective surface.
The two curves are rather close to
each other, suggesting that approximating
Eq.~(\ref{eq:approx1}) by Eq.~(\ref{eq:approx2}) is correct.
All curves have similar variations as in the Ca--Ar
diatomics,\cite{epjd} except a global shift to lower energies
due to the additional argon-argon interaction energies. The excited
state surfaces show a regular behavior, but we also note the
presence of a crossing near $d\sim 10\,a_0$ which is
also present in the Ca--Ar pair. The corresponding effective
absorption cross sections $\bar\sigma(d)$ are displayed in
Fig.~\ref{fig:int13}. Except at small distances $d<6\,a_0$,
they remain very close to their asymptotic atomic value.
Hence the
approximation $\overline{(AB)}\approx\bar A\times \bar B$ in
Eq.~(\ref{eq:approx1}) should be satisfied.

\begin{figure}[htb]
\vbox to 6.2cm{
\includegraphics{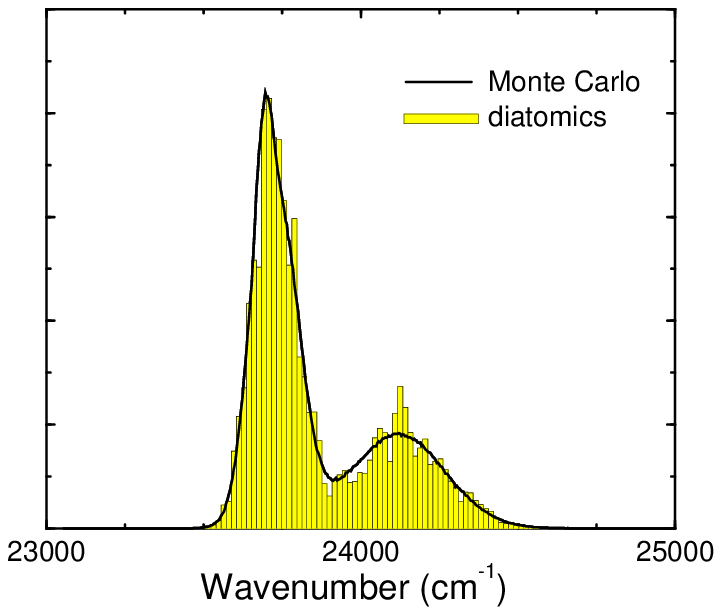}
\vfill}
\caption{Normalized simulated (solid line) and effective, diatomics-like
(histogram) absorption spectra of CaAr$_{13}$ at $T=30$~K, using the local
coordinate $d$.}
\label{fig:spec13}
\end{figure}

The photoabsorption spectrum calculated using the diatomics
picture and the effective potential surfaces is represented
as a normalized histogram in Fig.~\ref{fig:spec13}, along
with the result of classical Monte Carlo simulations. The
agreement is remarkable, as far as the positions and widths
of the peaks are concerned. The diatomics method introduces some additional
noise due the unaccuracy in the calculated effective
potential curves. This noise could be partly reduced by
taking a smaller number of bins in the interval of distances
$d$. However, this would also require one to reduce the
number of bins in the histogram of the absorption spectrum.

As compared to our previous study using the superposition
approximation,\cite{paperI} the absorption spectrum calculated using
classical Monte Carlo does not show a significant signature
of the isomers with calcium inside the icosahedral shell, the
extra peak found in Ref.~\onlinecite{paperI} being
replaced with a shoulder in the blue wing of the red peak.
This prevents a clear identification of the possible isomers
in the effective diatomics calculation.

\begin{figure}[htb]
\vbox to 6.2cm{
\includegraphics{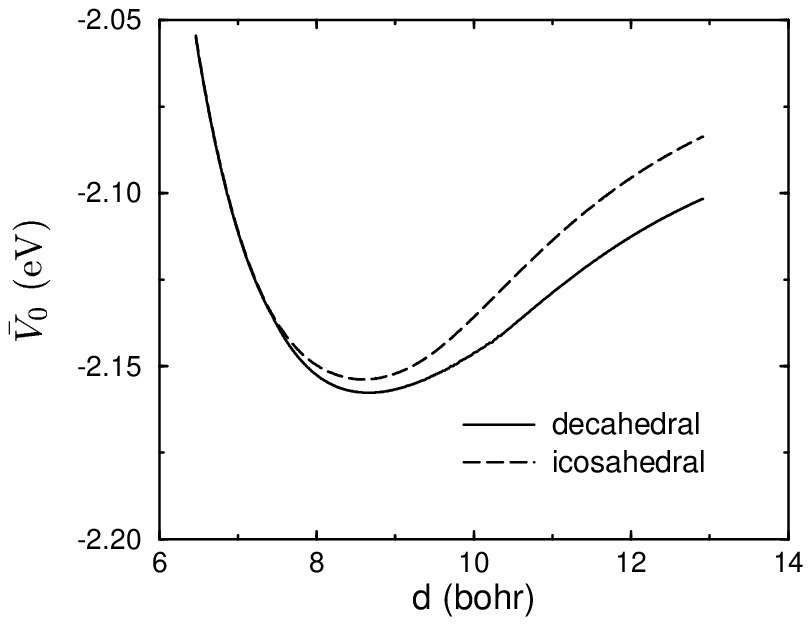}
\vfill}
\caption{Effective ground state potentials of CaAr$_{37}$ at $T=5$~K as
function of the local coordinate $d$, for the decahedral global minimum
(solid line) and the lowest Mackay icosahedral minimum (dashed line).}
\label{fig:pot37}
\end{figure}

\begin{figure}[htb]
\vbox to 7cm{
\includegraphics{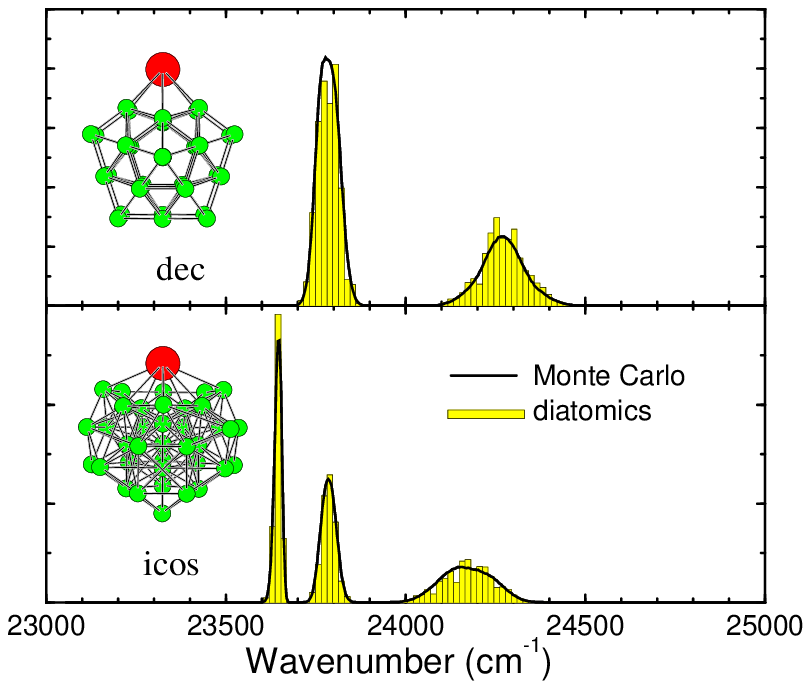}
\vfill}
\caption{Normalized simulated (solid line) and effective, diatomics-like
(histogram) absorption spectra of CaAr$_{37}$ at $T=5$~K, using the local
coordinate $d$.
Upper panel: decahedral minimum; lower panel: icosahedral minimum.}
\label{fig:spec37}
\end{figure}

The CaAr$_{37}$ cluster has a large number of stable
low-energy minima,\cite{paperI} characterized by very
different geometries. Its global minimum is decahedral, and
the next most stable isomer is a Mackay-type icosahedron. Due
to their different spectroscopic signatures at low
temperature, these two isomers provide a way of confronting
the diatomics method to more detailed spectroscopic data. In
Fig.~\ref{fig:pot37} we have represented the effective ground
state potential curves as a function of the local coordinate
$d$, for the two decahedral and icosahedral isomers, at
$T=5$~K. At such
low temperature, and using simple Monte Carlo, the cluster
is expected to be trapped in its initial basin.\cite{paperII}
As can be seen from Fig.~\ref{fig:pot37}, the icosahedral
structure remains more stable than the decahedral minimum.
The equilibrium distance is slightly shifted between the two
isomers, which results from the different local arrangements
of argon atoms near the calcium site. Neither the effective
excited state potential curves nor the effective absorption
cross sections are shown in this figure, as they display the
same variations as in CaAr$_{13}$ with the same reaction
coordinate. The normalized absorption spectra calculated with
the effective diatomics approach are compared in
Fig.~\ref{fig:spec37} to the reference Monte Carlo data.
Besides the presence of residual noise, we find again a good
agreement in the positions and widths of the absorption peaks
for the two isomers. At temperatures higher than 5~K,
multiple isomers become populated at thermal
equilibrium,\cite{paperI} but they are separated by large
energy barriers, making the sampling difficult even within
the Wang-Landau approach. In an experiment, such a situation would
correspond to the presence of several stable isomers in the
cluster beam. Turning back to calculations, rather than
trying algorithmic alteration to treat all the isomers in one
calculation, it seems more appropriate to perform calculations
for each isomer and then average calculations with proper
weights, in a spirit similar to the superposition approximation.\cite{paperI}

When the barrier between isomers is not as high as in
these pathological systems, isomerization is in principle
included in the calculated effective properties. However, the
choice of the local coordinate $d({\bf R})$ may not be
appropriate to reveal features associated with the presence
of several isomers. A more suitable reaction coordinate
would show distinct values depending on the isomers we want
to separate. There is indeed some arbitrariness in
the choice of $d({\bf R})$, which is very much guided by the
kind of process we want to focus on. In CaAr$_{13}$, the
distance $d_{\rm com}({\bf R})$ between the calcium atom and
the center of mass of the argon atoms has essentially two
values depending on the calcium atom being in a capping
location over the argon icosahedron or inside the icosahedral
shell. In the latter case, $d_{\rm com}$ can actually take 4
different values, one for each icosahedral site, but the 4
values are close to each other (around 10\,$a_0$),
and relatively far from the 13.5\,$a_0$ value in the
global minimum.

\begin{figure}[htb]
\vbox to 7.8cm{
\includegraphics{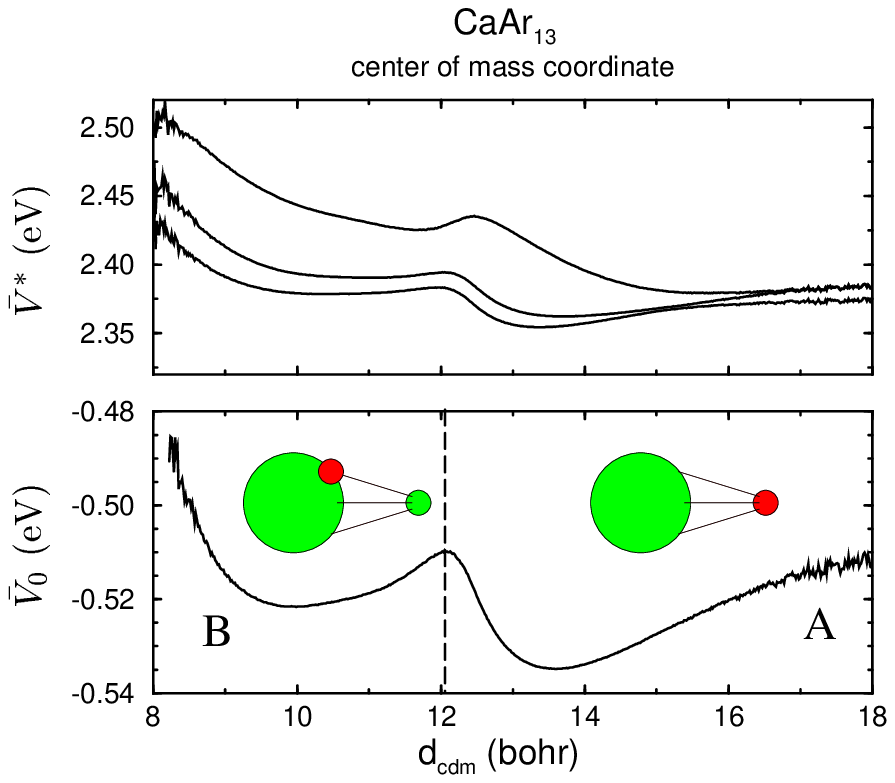}
\vfill}
\caption{Effective potential surfaces of CaAr$_{13}$ at $T=30$~K, as function
of the center-of-mass coordinate $d_{\rm com}$. $V_0$ and $V^*$ are the ground-
and excited-states surfaces, respectively. On the lower panel, the vertical
dashed line defines the two different structures of the cluster, where the
capping site is occupied either by calcium (A) or by argon (B).}
\label{fig:pot13cdm}
\end{figure}

The effective potential curves for CaAr$_{13}$ at $T=30$~K
using the $d_{\rm com}$ coordinate are represented in
Fig.~\ref{fig:pot13cdm}. We could not extend the sampling of
this coordinate below 8\,$a_0$ or beyond 18\,$a_0$ because of
extensive noise. The effective ground state potential clearly
shows the presence of two sets of minima, which can be
attributed to the calcium (A) or argon (B)
capping atom sketched in the lower panel of
Fig.~\ref{fig:pot13cdm}, depending on the outcome of a quenching procedure.
Regions (A) and (B) are defined by
the distance $d_{\rm com}$ being larger or smaller than
12.01$a_0$, respectively. The effective excited states
potential curves are also strongly influenced by this change
in coordinate, and their variations reflect the two stable
minima near 10$a_0$ and 13.5$a_0$, respectively. Therefore
the $d_{\rm com}$ reaction coordinate provides a structural
order parameter that can distinguish between the different
parts of the configuration space we are interested in. Being
able to differentiate isomers (A) and (B) allows one to
calculate the separate contributions of each region to the
global absorption spectrum. We have reported in
\begin{figure}[htb]
\vbox to 7.cm{
\includegraphics{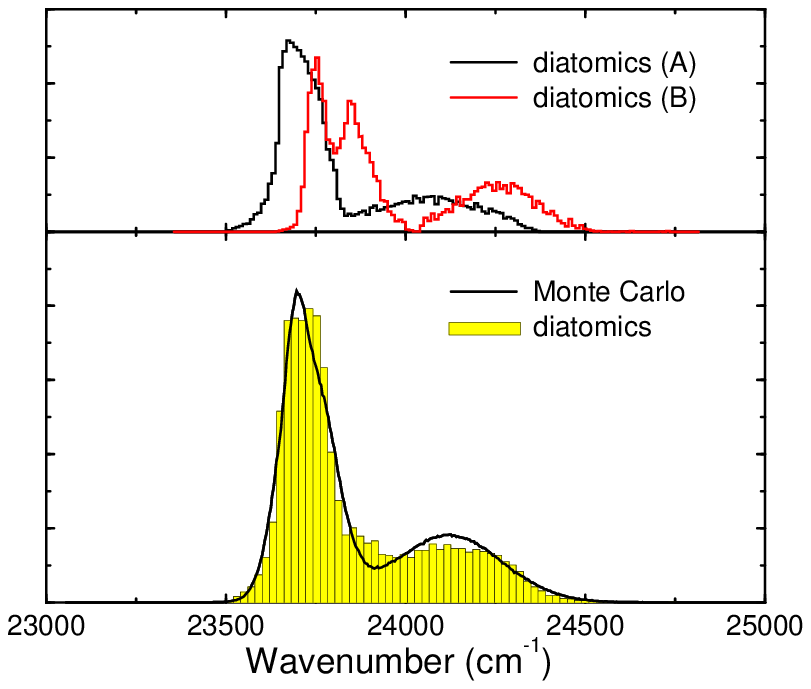}
\vfill}
\caption{Normalized simulated (solid line) and effective, diatomics-like
(histogram) absorption spectra of CaAr$_{13}$ at $T=30$~K, using the
center-of-mass coordinate $d_{\rm com}$.
On the upper panel, the normalized spectra
obtained by considering only regions A or B of the potential curves are
represented (see Fig.\protect~\ref{fig:pot13cdm}).}
\label{fig:spec13cdm}
\end{figure}
Fig.~\ref{fig:spec13cdm} the total absorption spectrum and
its individual contributions from (A) or (B). The overall
agreement between the global spectrum and the Monte Carlo
data is again good, and we notice that the blue shoulder
near 23900~cm$^{-1}$ is larger than when using the local
coordinate. This shows that the sampling of isomers (B) is
more efficient, maybe even slightly too efficient with
respect to the actual ergodic result. Longer simulations and
a better sampling of region (A) would be needed to reduce
this error. 

By carrying the integral (\ref{eq:approx2})
over each region (A) or (B) separetely allows for a
spectroscopic distinction between the isomers. It can be
seen that region (B) is responsible for the blue wing
at 23900~cm$^{-1}$. The global spectrum is the weighted sum
of the two separate contributions from (A) or (B), and the
statistical weight of (B) was found to be around 25\%, in
agreement with the study in Ref.~\onlinecite{paperI}. It is
interesting to discuss further the shape of the spectrum for
each isomer. We discussed in Ref.~\onlinecite{epjd} that the blue
and red bands of spectra, similar to that shown in
Fig.~\ref{fig:spec13cdm} for isomer A, are associated
to exciting the 4p orbital perpendicular or parallel to
the cluster surface, respectively. In the case of isomer A,
because of the outer location of calcium the two possible
parallel alignments of the 4p orbital are almost
degenerated in average and lead to two merging bands. In
contrast, calcium is closer to argon atoms in isomer B. As
seen in Fig.~\ref{fig:pot13cdm} this corresponds to a larger
the splitting between the two lower $(\hat V^*)$ potentials
curves in the region that
is accessible from the ground state isomer B. As a
result, the red band in Fig.~\ref{fig:spec13cdm} is splitted
into two components for isomer B. A splitted red band
has actually been observed experimentally 
in a slightly different context, the Ba$(6s^2\ ^1S_0 \to
6s6p\,^1P_1)$ excitation on large argon clusters.\cite{jmm94c}

\begin{figure}[htb]
\vbox to 7.6cm{
\includegraphics{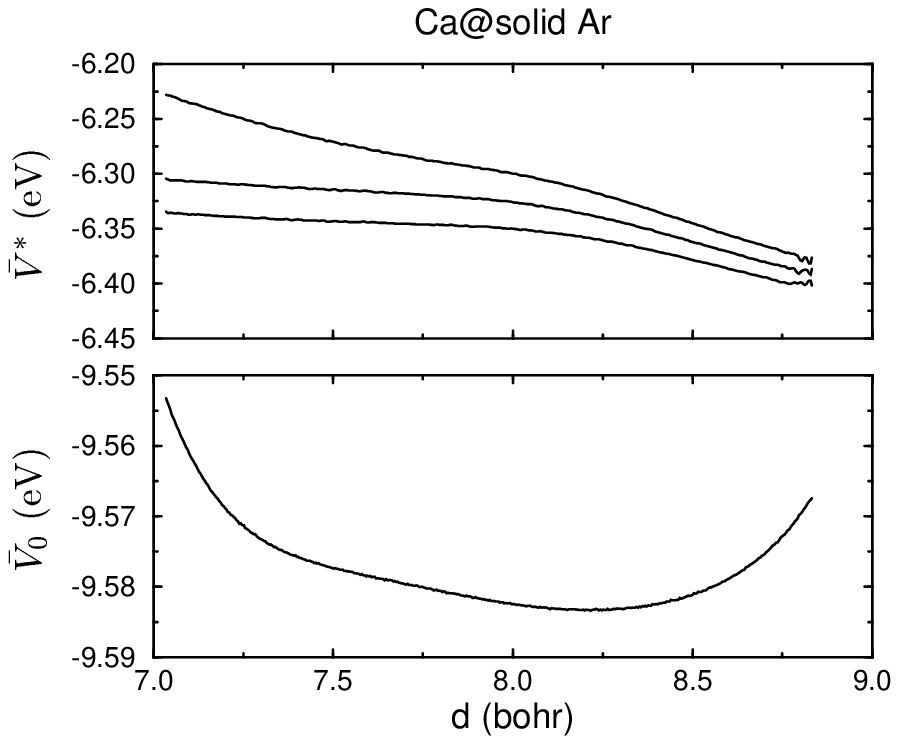}
\vfill}
\caption{Effective potential surfaces of Ca in solid argon at density
$\rho=1.562\times 10^3$~kg.m$^{-3}$ and temperature $T=20$~K, as
function of the local distance coordinate $d$. $V_0$ and $V^*$ are the ground-
and excited-states surfaces, respectively.}
\label{fig:potbulk}
\end{figure}

\begin{figure}[htb]
\vbox to 7cm{
\includegraphics{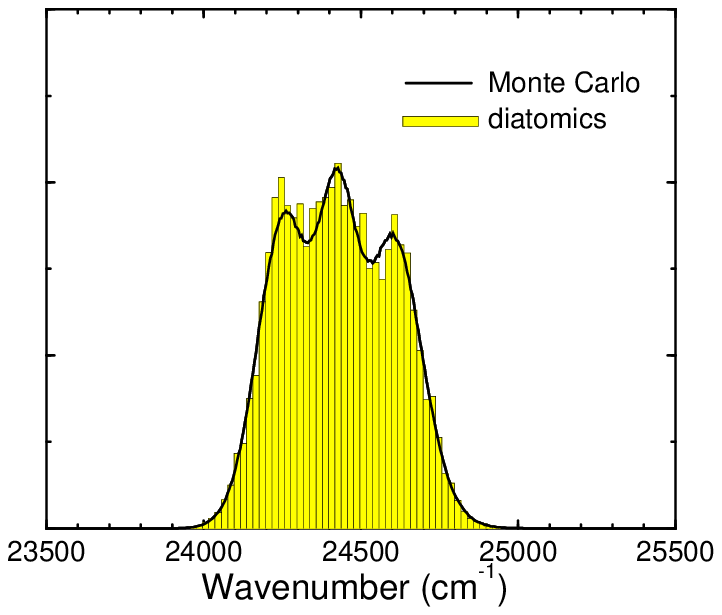}
\vfill}
\caption{Normalized simulated (solid line) and effective, diatomics-like
(histogram) absorption spectra of Ca in solid argon at density
$\rho=1.562\times 10^3$~kg.m$^{-3}$ and temperature $T=20$~K, using the
local coordinate $d$.}
\label{fig:specbulk}
\end{figure}

We just have seen that the spectroscopic properties
of the chromophore are substantially affected by partial
solvation in the argon cluster, that they appear as
sensitive probe of the local environment of calcium and that they
are satisfactorily described by the present effective diatomic
picture. Going further in that direction, we consider a
fully solvated Ca atom in an argon matrix as examined in the
Monte Carlo simulations of Ref.~\onlinecite{epjd}. The present
system is now a calcium atom surrounded by 107 argon
atoms in a face-centered cubic lattice, at constant density
$\rho=1.562\times 10^3$~kg.m$^{-3}$ and temperature
$T=20$~K. Bulk argon solvent is simulated using periodic
boundaries conditions in the minimum image convention.  The
reaction coordinate is again the local distance from calcium
to all other argon atoms, and the simulations were carried
out with the same statistics as for clusters. The effective
ground- and excited-states potential curves are displayed in
Fig.~\ref{fig:potbulk}. Since calcium remains fully solvated
in an equilibrium position, the ground-state surface roughly
has a distorted parabola shape, and cannot tend to zero at
large distances $d$. As we expect, the excited-states
surfaces are fully degenerated, giving rise to a dynamical
Jahn-Teller effect in absorption.\cite{epjd} The
photoabsorption spectrum calculated from the effective
diatomics approach is compared to Monte Carlo results in
Fig.~\ref{fig:specbulk}. The three excited potential
surfaces lead to three absorption peaks, but they are quite
broad, and the precise identification is less easy than with
the MC data. Still we observe a surprisingly good agreement
for the locations and overall widths of the peaks,
suggesting that the effective diatomics method is adequate
for treating chromophore-doped inert systems with general
shapes.

\section{Discussion and conclusion}
\label{sec:ccl}

Spectroscopy of polyatomic molecules with many degrees of
freedom is a difficult theoretical problem. Beyond simple
harmonic approximations,\cite{yan,wp} the most convenient
way to describe correctly the absorption intensity relies on
numerical simulations based on relevant Hamiltonians for the
potential energy surfaces. However the complex,
multidimensional character of the PES is a burden for simple
physical interpretations of the results obtained by
conventional methods. In this paper, we have presented a
simple alternative approach based on the separate treatment
of a single reaction coordinate, the many-body nature of the
problem being thermally averaged into effective
(free-)energy potential curves. By reducing the many-body
problem into a single coordinate problem, the system is
considered as a pseudo diatomics molecule, where the
interactions implicitely depend on the thermodynamical
conditions (temperature, but also pressure, and possibly
chemical potential) and include the fluctuations within the
averaged pseudo-atom. This point of view is best suited to the
case of CaAr$_n$ clusters, where the visible photoexcitation
is localized on the single calcium atom. In heterogeneous clusters,
extra reaction coordinates could naturally appear. For instance,
in CaKrAr$_n$, the two distances between calcium and krypton, and
between calcium and the argon cluster could be treated on a same
footing. In such cases, the present investigation could be easily
generalized to provide effective hypersurfaces depending on several
coordinates. In this picture the example chosen above would be
considered as a pseudo triatomics Ca--Kr--Ar$_n$.
Similar ideas could be applied to atom-diatom reactions
solvated by an inert cluster.

Computationally speaking, the present method involves the
preliminary calculation of the potential of mean force,
which incorporates the statistical averages over all but one
coordinates. The PMF is then used in a reweighting
simulation to obtain the effective interactions and
absorption cross section in a broad range of distances. The
overall numerical cost is therefore heavier than in a
conventional simulation, by one order of magnitude at the
very least.  However, the calculation of the PMF can be
achieved conveniently, independently of the reaction
coordinate chosen, using the multiple range random walk
algorithm of Wang and Landau,\cite{wl} recently adapted to
the problem of free-energy profiles.\cite{fcmp} Therefore
the interest of the present approach mainly resides in the
extra interpretations it provides. For instance, it could be
used to assess or quantify the local character of the
excitation. It also lays some ground for further
spectroscopic investigations. In particular, more detailed
studies of the excited states effective surfaces could be
undertaken. The influence of temperature or the number of
inert atoms on these curves, the related possible conical
intersections offer examples of future research directions.

On a practical point, we have largely used a local distance
coordinate $d$, namely the geometric distance ({\em i.e.}
the lowest distance) between calcium and argon atoms. This
order parameter is relevant to describe the CaAr$_n$ cluster
as a pseudo Ca--(Ar$_n$) diatomics, for which most of the
methodology developed here was meant. However, the separate
effects of different isomers are essentially smoothed out
and averaged, and it may be much more profitable to use
other system-dependent coordinates to get a richer
information about structure. In the case of CaAr$_{13}$, we
have shown that the distance between calcium and the argon
cluster center of mass was more appropriate in this
purpose. In other, more complex clusters, one should take
profit of the known structural information to design
suitable reaction coordinates. For instance the bond order
parameters $Q_4$, $Q_6$, $W_4$, or $W_6$, introduced by
Steinhardt and coworkers,\cite{bop} could be used to
distinguish between icosahedral, octahedral, or decahedral
isomers.\cite{ar38doye,ar38fc,ar75doye} One could then
extend the present work to metallic clusters, where the energy
levels can change upon global shape deformation,\cite{landman}
and for which appropriate reaction coordinates could be the
eccentricities or the Hill-Wheeler parameters.

Quantum vibrational effects are not accounted for in the
present description, essentially because the corresponding
treatment within path-integral finite temperature Monte
Carlo is numerically demanding, and because reference
calculations are not available. However, at moderate
temperatures, the present approach could be
straightforwardly applied to quantum-corrected potential
energy surfaces using the Feynman-Hibbs effective
potentials.\cite{fh}

At a more general level, the present approach gives
the framework both to describe a reactive system in terms of
deformations along a reaction coordinate and to calculate
the corresponding  energy variation. In that sense, the
present calculation allows one to deal with the description
of a chemical reaction in terms of the transition state
theory in situations where part of the reactive system
itself acts as a thermal bath. In the present case the
reactive system is a calcium atom plus an argon cluster. The latter
plays the role of the thermal bath. The reaction is the
simplest that might be considered: solvation of Ca by the
argon cluster. The choice of reaction coordinate is not
unique and a possible choice is
the distance between Ca and the closest argon atom. Finally
the relevant energy term when the system is free to evolve
along the reaction coordinate is the potential of mean force 
that was define in the present work as a
Helmoltz free energy term.

\end{document}